\documentstyle{mn}
\input epsf

\newif\ifAMStwofonts

\ifoldfss
  \ifCUPmtlplainloaded \else
    \NewTextAlphabet{textbfit} {cmbxti10} {}
    \NewTextAlphabet{textbfss} {cmssbx10} {}
    \NewMathAlphabet{mathbfit} {cmbxti10} {} 
    \NewMathAlphabet{mathbfss} {cmssbx10} {} 
  \fi
  \ifAMStwofonts
    \ifCUPmtlplainloaded \else
      \NewSymbolFont{upmath} {eurm10}
      \NewSymbolFont{AMSa} {msam10}
      \NewMathSymbol{\upi}     {0}{upmath}{19}
      \NewMathSymbol{\umu}     {0}{upmath}{16}
      \NewMathSymbol{\upartial}{0}{upmath}{40}
      \NewMathSymbol{\leqslant}{3}{AMSa}{36}
      \NewMathSymbol{\geqslant}{3}{AMSa}{3E}

      \let\leq=\leqslant 
      \let\geq=\geqslant 
    \fi
  \fi
\fi 

\ifnfssone
  \newmathalphabet{\mathit}
  \addtoversion{normal}{\mathit}{cmr}{m}{it}
  \addtoversion{bold}{\mathit}{cmr}{bx}{it}
  \newmathalphabet{\mathbfit} 
  \addtoversion{normal}{\mathbfit}{cmr}{bx}{it}
  \addtoversion{bold}{\mathbfit}{cmr}{bx}{it}
  \newmathalphabet{\mathbfss} 
  \addtoversion{normal}{\mathbfss}{cmss}{bx}{n}
  \addtoversion{bold}{\mathbfss}{cmss}{bx}{n}
  \ifAMStwofonts
    \ifCUPmtlplainloaded \else
      \UseAMStwoboldmath
      \makeatletter
      \new@mathgroup\upmath@group
      \define@mathgroup\mv@normal\upmath@group{eur}{m}{n}
      \define@mathgroup\mv@bold\upmath@group{eur}{b}{n}
      \edef\UPM{\hexnumber\upmath@group}
      \new@mathgroup\amsa@group
      \define@mathgroup\mv@normal\amsa@group{msa}{m}{n}
      \define@mathgroup\mv@bold\amsa@group{msa}{m}{n}
      \edef\AMSa{\hexnumber\amsa@group}
      \makeatother
      \mathchardef\upi="0\UPM19
      \mathchardef\umu="0\UPM16
      \mathchardef\upartial="0\UPM40
      \mathchardef\leqslant="3\AMSa36
      \mathchardef\geqslant="3\AMSa3E

      \let\leq=\leqslant 
      \let\geq=\geqslant 
    \fi
  \fi
\fi 

\ifnfsstwo
  \DeclareMathAlphabet{\mathbfit}{OT1}{cmr}{bx}{it}
  \SetMathAlphabet\mathbfit{bold}{OT1}{cmr}{bx}{it}
  \DeclareMathAlphabet{\mathbfss}{OT1}{cmss}{bx}{n}
  \SetMathAlphabet\mathbfss{bold}{OT1}{cmss}{bx}{n}

  \ifAMStwofonts
    \ifCUPmtlplainloaded \else
      \DeclareSymbolFont{UPM}{U}{eur}{m}{n}
      \SetSymbolFont{UPM}{bold}{U}{eur}{b}{n}
      \DeclareSymbolFont{AMSa}{U}{msa}{m}{n}
      \DeclareMathSymbol{\upi}{0}{UPM}{"19}
      \DeclareMathSymbol{\umu}{0}{UPM}{"16}
      \DeclareMathSymbol{\upartial}{0}{UPM}{"40}
      \DeclareMathSymbol{\leqslant}{3}{AMSa}{"36}
      \DeclareMathSymbol{\geqslant}{3}{AMSa}{"3E}

      \let\leq=\leqslant 
      \let\geq=\geqslant 
    \fi
  \fi
\fi 

\ifCUPmtlplainloaded \else
  \ifAMStwofonts \else 
    \def\upi{\pi}
    \def\umu{\mu}
    \def\upartial{\partial}
  \fi
\fi

\title[Dynamical friction in a gaseous sphere]
  {Dynamical friction of bodies orbiting in a gaseous sphere}
\author[F. J. S\'{a}nchez-Salcedo and A. Brandenburg]
  {F. J.~S\'{a}nchez-Salcedo\thanks{Present address:
  Instituto de Astronomia, UNAM, Ap. 70 264, Mexico D. F. CP 04510;
Email: jsanchez@astroscu.unam.mx} and
  A.~Brandenburg\thanks{Also at: Nordita, Blegdamsvej 17,
  DK-2100 Copenhagen \O, Denmark; Email: brandenb@nordita.dk}\\
  Department of Mathematics, University of Newcastle, Newcastle upon Tyne, NE1 7RU}
\date{Accepted 2000 September 29. Received 2000 September 15;
in original form 1999 December 22}

\pagerange{\pageref{firstpage}--\pageref{lastpage}}
\pubyear{...}

\def\LaTeX{L\kern-.36em\raise.3ex\hbox{a}\kern-.15em
    T\kern-.1667em\lower.7ex\hbox{E}\kern-.125emX}

\begin{document}

\label{firstpage}

\maketitle
\begin{abstract}
The dynamical friction experienced by a body moving in a gaseous medium
is different from the friction in the case of a collisionless
stellar system. Here we
consider the orbital evolution of a gravitational perturber inside a
gaseous sphere using three-dimensional simulations, ignoring however
self-gravity. The results are analysed in terms of a `local' formula
with the associated Coulomb logarithm taken as a free parameter.
For forced circular orbits, the asymptotic value of the component of the
drag force in the direction of the velocity is a slowly varying function
of the Mach number in the range $1.0$--$1.6$.
The dynamical
friction timescale for free decay orbits is typically only half as long as 
in the case of a collisionless background, which is in agreement with
E.C.~Ostriker's recent analytic result.  The orbital decay rate
is rather insensitive to the past history of the perturber.  It is
shown that, similar to the case of stellar systems,
orbits are not subject to any significant circularization.
However, the dynamical friction timescales
are found to increase with increasing orbital eccentricity for the
Plummer model, whilst no strong dependence on the initial eccentricity
is found for the isothermal sphere.
\end{abstract}


\section{Introduction}
A gravitating body moving through a background medium suffers from a drag force
due to the interaction with its own induced wake. Generally speaking, the medium
may be composed of collisionless matter, and/or gas. For collisionless backgrounds, 
the classical Chandrasekhar formula has
proved useful in determining the orbital decay of galactic satellites
and globular clusters around spherical systems such as elliptical galaxies (e.g. Lin \&
Tremaine 1983; Cora, Muzzio \& Vergne 1997, and references therein), even though
it was inferred for uniform media.
For a uniform and infinite gaseous medium, the gravitational drag on a
body moving at constant velocity on a straight-line orbit, has been estimated  
both by linear theory (e.g. Ruderman \& Spiegel 1971; Rephaeli \& Salpeter 1980;
Ostriker 1999) and from numerical experiments in different settings
(e.g. Shima et al.\ 1985; Shima et al.\ 1986; Shankar, Kley \& Burkert 1993;
Kley, Shankar \& Burkert 1995; Ruffert 1996).
All these authors consider a uniform medium because they are
mainly interested in the accretion flow past the body.
Here we will investigate
the gaseous drag and the sinking decay of a light body orbiting on a 
gaseous sphere which is initially in hydrostatic equilibrium with a given
gravitational potential. 
In fact, dynamical friction for gaseous and spherical backgrounds has
its relevance to many astrophysical studies, e.g. 
galaxies in galaxy clusters, the relaxation of young stellar clusters,
or the dynamics of massive black holes within their host galaxies.
It will be useful to check in which conditions
the drag formula obtained in linear theory for uniform media works for spherical
backgrounds, in analogy to stellar systems.

The question of dynamical friction for gaseous and spherical backgrounds has received
new interest in connection with recent studies suggesting that Galactic
dark matter may be in the form of cold molecular clouds either
distributed in a disc \cite{Pfe94} or in a quasi-spherical halo (e.g.
De Paolis et al.\ 1995; Gerhard \& Silk 1996; Walker \& Wardle 1998;
Walker 1999; De Paolis et al.\ 1999; Sciama 2000).  For certain
parameters of these clouds, the assumption of collisionless matter is
no longer valid. Other authors have proposed, based on
observational astrophysical grounds, that dark matter may
be self-interacting with a large scattering cross-section
\cite{Spe00}. It is therefore important to find out the dynamical
implications of having spherical halos of collisional matter 
and, in particular, its effect on the decay of satellite galaxy
orbits.

Deeper physical insight into the
question of how dynamical friction is affected by collisions may be
provided by considering continuous gaseous media.  
Recently,
Ostriker \shortcite{Ost99} pointed out that for supersonically
moving bodies the drag in a uniform gaseous medium is more efficient
than in the case of collisionless media that are described by the
standard Chandrasekhar formula; and is non-vanishing
for subsonic bodies. The same feature is observed
in the case of a perfectly absorbing body \cite{Ruf96}. 
As a consequence, satellite galaxies
may experience more rapid decay in halos made up of molecular
clouds, especially in earlier epochs before most of the gas
has turned into stars.

In the standard model of cosmological structure formation, forming
substructures continuously exchange orbital energy with both
collisionless material and with their surrounding hot medium due to
dynamical friction. An understanding of the processes that can produce
velocity and spatial bias where the concentration and velocity
dispersion is different for different galactic populations, e.g. mass
segregation in clusters of galaxies, is essential for the
interpretation of observational data as well as for semi-analytical
models of the formation of galaxies and galaxy clusters. Clumps of
cold gas moving through a hotter medium may suffer a hydrodynamical
drag that is stronger than the gravitational drag (e.g.\ Tittley et
al.\ 1999). In this work, however, we focus on a purely gravitational
perturber.

The linear response of a spherical system to an orbiting gravitational perturber
was carried out by Balbus \& Soker \shortcite{Bal90} motivated by the
problem of excitation of internal gravity waves by galaxies in the core
of clusters.  In this paper we concentrate on the dynamical friction of
a rigid perturber orbiting in a gaseous medium that shows a spherical
density distribution with a concentration towards the centre.  
Initially the gas is at rest and in hydrostatic equilibrium with an external
spherical gravitational potential.
The gravitational perturber is assumed to be very large (typically
the softening gravitational radius is taken a few times the 
accretion radius) and
unable to accrete matter. We do not model the existence of a
physical surface on the body, i.e. no inner boundary condition on the
body has been imposed. A discussion on the importance
of accretion is given in Section 2.  
Our aim is to compute the evolution and sinking
time of the perturber towards the center. 

The assumptions of infinity and homogeneity of the medium, that was
required in Ostriker's analysis, are relaxed, and, in addition, the
contribution of non-local effects on the drag force are taken into
account. Also, an additional difference is that we deal with extended
perturbers whose radius is a significant fraction of the size of the
background system.  Since the perturbers in most cases of interest do
not present circular orbits \cite{Bos99}, and also motivated by the
fact that cosmological simulations show that the majority of satellite
orbits have quite large eccentricities, we consider the dependence of
the sinking times on the eccentricity of the orbits.

There are some similarities with the dynamical evolution of binary
star cores embedded in a common envelope (e.g. Taam, Bodenheimer \& R\'{o}zyczka 1994; and
references therein). In those systems, however, the gravitational torques
of the two cores are able to spin up the envelope to a high rotation
(see also Sandquist et al.\ 1998). The
fact that the masses of the stars and the envelope are comparable is crucial for the subsequent
evolution of the system. Here we explore a rather different case in which there 
is a single rotating body and its mass is significantly less than that
in the gaseous background mass.
Recall that neither wind is injected in the gaseous
component nor radiation is considered in the present work. 

The paper is organized as follows.
In Section 2, we highlight the differences between dynamical friction
in collisionless and collisional media; the importance of having 
mass accretion by the perturber is also addressed. 
The description of the model is given
in Section 3. In Section 4 the orbital decay rate
is computed numerically for a body orbiting around a gaseous sphere, and is
compared with a `local' estimate of the drag force.
Implications and conclusions are given in Section 5.

\section{The effects of collisions and accretion on dynamical friction}

\subsection{Comparing the drag force in collisionless and collisional
backgrounds: a statistical discussion}

Dynamical friction may be pictured as the response of
the system which tends to equipartition.
The most classical problem is the motion of a `macroscopic'
particle travelling in a fluid. The evolution of such
a particle is described by the Fokker-Planck
equation with the diffusion tensor depending on the correlations
of the fluctuating force in the unperturbed state (e.g. Isihara 1971).
These fluctuations in the force are generated as a consequence of the
graininess of the background, which may be either collisional or
collisionless.

For collisionless systems, Chandrasekhar \& von Neumann
\shortcite{Cha42}, Lee \shortcite{Lee68}, Kandrup \shortcite{Kan83},
Bekenstein \& Maoz \shortcite{Bek92} and Colpi \shortcite{Col98}, among
others, computed the autocorrelation tensor in a stellar system by
making the approximation that each background particle follows a linear
trajectory with constant velocity. The inclusion of curved orbits does
not change the result appreciably \cite{Lee68}. Assuming a Maxwellian
distribution of field particle velocities, Kandrup \shortcite{Kan83}
and Bekenstein \& Maoz \shortcite{Bek92} derived a formula identical to
the Chandrasekhar formula from the correlation tensor.

For collisional systems, it is usually assumed that the mutual
encounters between field particles destroy the correlation of each
field particle's orbit with itself. However, from this it does not
follow that scattering of field particles reduces the correlation
tensor and so the dynamical friction. For instance, Ostriker
\shortcite{Ost99} considered the dissipative drag force experienced by
a perturber of mass $M_{\rm p}$ with velocity $\bmath{V}$ moving
through a homogeneous gaseous medium of sound speed $c_{\rm s}$ and
unperturbed density $\rho_{0}$.  Both the linearized Euler equations
\cite{Ost99} and numerical calculations (S\'{a}nchez-Salcedo \&
Brandenburg 1999, hereafter referred to as Paper I) show that the
gravitational drag in such a fully collisional medium exceeds the value
given by the Chandrasekhar formula, provided that the Mach number,
${\mathcal{M}}\equiv V/c_{\rm s}$, lies in the range $1$--$2.5$.
\footnote{For the drag force in the case of an absorbing perturber we refer
to the paper by Ruffert (1996).}

In order to ease visualization it is convenient to consider the gaseous
medium as a system of a large number of interacting field particles of
identical masses $m$. In the hydrodynamical limit the distribution of
velocities is always Maxwellian (which is not true for collisionless
systems). Let us assume that in such a limit the interaction between
field particles is large enough that the {\it effective} velocity in
each direction is close to $c_{\rm s}$ for the field particles. An {\it
estimate} of the force in gaseous media emerges immediately by
extending Chandrasekhar's results. For collisionless backgrounds,
Chandrasekhar (1943) showed that ambient particles with speeds lower
than the object contribute to the drag force as
\begin{equation}
F_{\rm df}=\frac{4\pi G^{2} M^{2}_{\rm p} m}{V^{2}} \int_{0}^{V} dv\,f(v)
\left(\ln \Lambda +\ln \frac{V^{2}-v^{2}}{V^{2}}\right),
\label{eq:Ch1}
\end{equation}
whereas field particles moving faster than the perturber contribute
with the lower-order term
\begin{equation}
F_{\rm df}=\frac{4\pi G^{2} M^{2}_{\rm p} m}{V^{2}}\int_{V}^{\infty} dv\,f(v)
\left[\ln\left(\frac{v+V}{v-V}\right)-\frac{2V}{v}\right].
\label{eq:Ch2}
\end{equation}
In the equations above $f(v)\,dv$ is the number of field particles with velocities
between $v$ and $v+dv$, and
$\Lambda\equiv b_{\rm max}/b_{\rm min}$, with $b_{\rm max}$ the characteristic
size of the medium and $b_{\rm min}={\rm max}\{r_{\rm min},GM_{\rm p}/V^{2}\}$, where
$r_{\rm min}$ is the characteristic size of the perturber.
Now, if we blindly use the above equations for a background of particles
$f(v)=N_{0}\delta(v-c_{\rm s})$, with $N_{0}=\rho_{0}/m$, and keep only the dominant term
for $V>c_{\rm s}$, we get
\begin{equation}
F_{\rm df}=\lambda_{1} \frac{4\pi G^{2} M^{2}_{\rm p}
\rho_{0}}{V^{2}}\ln \left(\frac{Vt}{b_{\rm min}}\right)
+{\mathcal{O}} \left(\ln\left[1-{\mathcal{M}}^{-2}\right]\right),
\label{eq:Ch1ii}
\end{equation}
where again ${\mathcal{M}}=V/c_{\rm s}$ and $b_{\rm max}\approx Vt$ was used. A discussion about
this dependence of $b_{\rm max}$ can be found in Ostriker \& Davidsen \shortcite{Ost68}.
For $V<c_{\rm s}$ we have
\begin{equation}
F_{\rm df}=\lambda_{2} \frac{4\pi G^{2} M^{2}_{\rm p} \rho_{0}}{V^{2}}
\left[\ln\left(\frac{1+{\mathcal{M}}}{1-{\mathcal{M}}}\right) - 2{\mathcal{M}}\right],
\label{eq:Ch2ii}
\end{equation}
where $\lambda_{1}$ and $\lambda_{2}$ are dimensionless parameters of order
unity. The first result is that the drag force is nonvanishing even for subsonic
perturbers, as occurs in collisionless backgrounds. We note that the
lower-order terms in Eq.~({\ref{eq:Ch2}) are often ignored in the
literature, which has sometimes led to wrong premises. For instance,
Zamir (1992) ignored such lower-order terms and
concluded that an object launched in a homogeneous and isotropic
background with a cutoff in the velocity distribution from below
could even be accelerated.

We may compare the above estimates with the exact values obtained by computing
the gravitational wake behind the body. Ostriker \shortcite{Ost99} showed in linear theory
that the drag force is given by:
\begin{equation}
F_{\rm df}=\frac{4\pi G^{2}M^{2}_{\rm p}\rho_{0}}{V^{2}}
\left[\ln\left(\frac{Vt}{r_{\rm min}}\right)+
\frac{1}{2}\ln \left(1-{\mathcal{M}}^{-2}\right) \right],
\label{eq:Ost1}
\end{equation}
for ${\mathcal{M}}\equiv V/c_{\rm s}>1$ and $t>r_{\rm min}/(V-c_{\rm s})$, and
\begin{equation}
F_{\rm df}=\frac{4\pi G^{2}M^{2}_{\rm p}\rho_{0}}{V^{2}}
\left[\frac{1}{2}\ln\left(\frac{1+{\mathcal{M}}}{1-{\mathcal{M}}}\right) -{\mathcal{M}}\right],
\label{eq:Ost2}
\end{equation}
for ${\mathcal{M}}<1$ and $t>r_{\rm min}/(c_{\rm s}-V)$.  It was
assumed that the perturber is formed at $t=0$. A minimum radius $r_{\rm
min}$ was adopted in order to regularise the gravitational potential of
a point mass.  We see that the expressions (\ref{eq:Ch1ii}) and
(\ref{eq:Ch2ii}) match Eqs (\ref{eq:Ost1}) and (\ref{eq:Ost2}),
respectively, if we take $\lambda_{1}=1$ and $\lambda_{2}= 1/2$
\footnote {Note that the lower-order term of equation (\ref{eq:Ch1ii}),
which is of order ${\mathcal{O}}(\ln^{-1}\Lambda)$, must have a
coefficient $1/2$ to match Eq.~(\ref{eq:Ost1}).}. This suggests
that many features of the dynamical friction for stellar systems, such
as the role of self-gravity or the importance of non-local effects, may
be shared by gaseous systems.

In the next section we analyse the problem of orbital decay in a
spherical and gaseous system. For stellar systems it was shown that
Chandrasekhar's formula is a good approximation (e.g., Lin \& Tremaine
1983; Cora et al.\ 1997). It is interesting to check whether Ostriker's
formula works also in such geometry, in analogy to stellar systems.

\subsection{Comparison with a totally absorbing perturber}

Considerable work has been devoted to studying the
hydrodynamics of the Bondi-Hoyle-Lyttleton
(hereafter BHL) accretion problem,
in which a totally
absorbing body interacts gravitationally with
the surrounding gaseous medium. In this case two different forces acting
upon the accretor can be distinguished: the gravitational drag caused by
the asymmetry distribution of mass, and the force associated with the
accretion of momentum by the perturber.
This case has been considered numerically by Ruffert (1996). 
Although the case of a totally absorbing accretor is
different from the non-absorbing one considered in
the present paper, it is interesting to compare
the two cases.

Ruffert \shortcite{Ruf96} showed that there is a gravitational drag
on a totally absorbing body that saturates very quickly in the subsonic regime,
and is roughly one order of magnitude smaller than in the supersonic models.
This is qualitatively similar to the case without accretion, as described by
Eqs (\ref{eq:Ost1}) and (\ref{eq:Ost2}), although the exact strength
of the gravitational drag may be somewhat different.
At first glance,
the gravitational drag seems to saturate or even decline
with time in Ruffert's supersonic experiments. By contrast,
in both Ostriker's analysis and
in our numerical simulations with no mass accretion
the gravitational drag clearly increases logarithmically in time.

Ruffert focuses mainly on the case $R_{\rm soft}\ll R_{\rm ac}$,
where $R_{\rm ac}\equiv2GM_{\rm p}/V^{2}$ is the accretion radius.
Whilst that case is relevant to BHL accretors, we are mainly interested in the
opposite case of small accretion radii.
In order to facilitate comparison, we will compare
the values of the gravitational drag found by Ruffert \shortcite{Ruf96}
for his largest accretor models, EL, FL, GL and HL,
where the size of the accretor is one accretion radius,
with the values given by Eqs (\ref{eq:Ost1}) and (\ref{eq:Ost2}).

For large accretors, the hydrodynamical force due to
accretion acts also as a drag force (e.g.\ Ruffert 1996).
That situation corresponds to
geometrical accretion of mass which tends
to suppress the flux of mass from
the forward side of the accretor to the back side
in the {\it supersonic} case.
This fact smears out the asymmetry of the mass density
distribution between both sides and, consequently,
the gravitational drag is expected to be reduced.

Ruffert (1996) gives the evolution of the gravitational drag force in
units where $R_{\rm ac}=c_{\rm s}=\rho_{0}=1$. In order to obtain the
result in units of $4\pi G^{2}M_{\rm p}^{2}\rho_{0}/c_{\rm s}^{2}$, we have
to divide Ruffert's values by $\pi{\mathcal M}^4$. Thus, for his model
GL with ${\mathcal M}=3$ the value 80 at time $t=50 R_{\rm ac}/c_{\rm s}$
(see his Fig.~5b) corresponds to $0.31$. For his model
HL with ${\mathcal M}=10$ the value 800 at time $t=15 R_{\rm ac}/c_{\rm s}$
(see his Fig.~7b) corresponds to $0.025$.
These values are somewhat smaller than those obtained
by using $r_{\rm min}$ approximately the size of the accretor
in Eq. (\ref{eq:Ost1}),
i.e. $0.55$ and $0.045$, respectively. 
In order to trace the origin of these discrepancies
we have checked in our records the
drag force for ${\mathcal M}=3$ and
$R_{\rm soft}=R_{\rm ac}$ (a resolution of $400\times 800$
zones was used, with a mesh width at the body position
of $1/8$ accretion radius). At time 
$t=50 R_{\rm ac}/c_{\rm s}$ we obtained a 
dimensionless drag force of $0.37$. The remaining $20\%$ discrepancy
may be associated with the fact that Ruffert's perturbers are absorbing.
This confirms our expectations that the gravitational drag 
should be smaller for an absorbing perturber, but note
that the total drag (gravitational plus momentum accretion
drag) is stronger.

Larger differences, however, are apparent for subsonic
motions. In fact, the gravitational drag for a body travelling with
Mach number 0.6 (model EL; see Fig.~2b in Ruffert \shortcite{Ruf96}) is
a factor $4$--$5$ {\it larger} than predicted by Eq.~(\ref{eq:Ost2}).
This may be a consequence of the fact that, in the {\it subsonic} case,
accretion is more efficient in loading down mass from the forward side
than from the back side.
Therefore we conclude that, for large accretors, mass accretion
would be important for dynamical friction in the subsonic regime.
After this short digression we now return, however, to the case of non-absorbing
perturbers with small accretion radii.

\section{Setup of the model}
We have performed several 3D simulations to study the dynamical
friction effect on a rigid perturber (or satellite) orbiting within a
partly gaseous and initially spherical system, which is modelling the
primary or central galaxy, with gas density $\rho(\bmath{r},t)$. 
The equation of state is that of a perfect gas with specific heat
ratio $\gamma=5/3$. Firstly, we
will assume that the gas is polytropic,
$P\propto \rho^{\Gamma}$, where the polytropic index
$\Gamma$ ranges between $1$ and $5/3\equiv\gamma$, the limits corresponding to the
gas being isothermal and adiabatic, respectively.  The perturber is
modelled by a rigid Plummer model with the corresponding gravitational
potential $\phi_{\rm p}=-GM_{\rm p}/\sqrt{r^{2}+R_{\rm soft}^{2}}$,
where $G$ is the gravitational constant.  In the absence of the
perturber, the gas is initially in hydrostatic equilibrium in the
overall potential given by $\phi_{\ast}+\phi_{\rm g}$, where $\phi_{\rm
g}$ is the potential created by the gas component and $\phi_{\ast}$ by
the rest of the mass. The potential $\phi_{\ast}$ is assumed to be
fixed in time, and the centre of the primary fixed in space.
Self-gravity of the gas is ignored which seems to be a reasonable
approximation to reality (e.g. Prugniel \& Combes 1992). For simplicity
we assume that the combined potential $\phi_{\rm
G}=\phi_{\ast}+\phi_{\rm g}$ is also given by a Plummer potential with
softening radius $R_{\rm G}$ and total mass $M_{\rm G}$.  As a
consequence, a polytropic system of pure gas must have polytropic index
$\Gamma=1.2$ in order to have a self-gravitating model (e.g. Binney \&
Tremaine 1987). For $\Gamma > 1.2$ the models above are not
self-consistent at large radii because the gravitational force due to
the gas increases faster than $|\bmath{\nabla}\phi_{\rm G}|$. However,
we use simple initial condition since our aim is to test the
applicability of various fit formulae which should be able to predict
the evolution of the satellite regardless of the full consistency of
the background system.  
For completeness, the case of $\phi_{\rm G}$ being
a King model instead of a Plummer model is also explored in order to
mimic the inner portions of a self-gravitating sphere; see Section
\ref{sec:sppis}.

Using an explicit cartesian code that is sixth order in space and third
order
in time, we solve numerically the continuity and Euler equations which,
for a polytropic gas, are
\begin{equation}
\frac{\partial\rho}{\partial t}+\bmath{\nabla}\cdot(\rho\bmath{v})=0,
\label{eq:fluid1}
\end{equation}
\begin{equation}
\frac{\partial \bmath{v}}{\partial t}+(\bmath{v}\cdot
\bmath{\nabla})\bmath{v}=
-\bmath{\nabla}(h+\phi_{\rm G}+\phi_{\rm p})
+\mbox{viscous force},
\label{eq:fluid2}
\end{equation}
where $h$ is related to the specific enthalpy and is defined by
\begin{equation}
h(\rho)=\int_{\rho_0}^{\rho} \frac{dP}{\rho}=
\left\{ \begin{array}{ll} \frac{\Gamma c_{{\rm s}0}^{2}}{\Gamma-1}
\left[\left(\frac{\rho}{\rho_0}\right)^{\Gamma-1}-1\right]
& \mbox{${\rm if} \;\; \Gamma\neq 1$}\\
c_{{\rm s}0}^{2} \ln(\rho/\rho_0)
 & \mbox{${\rm if} \;\; \Gamma=1$}
\end{array}
\right.,
\end{equation}
where $c_{{\rm s}0}=\sqrt{P_{0}/\rho_0}$,
$P_{0}$ and $\rho_{0}$ are the reference values of pressure and density,
respectively,
and
\begin{equation}
\phi_{\rm p}=-\frac{GM_{\rm p}}{\sqrt{(\bmath{r}-\bmath{R}_{\rm
p}(t))^{2}+R_{\rm soft}^{2}}}
\end{equation}
is the gravitational potential
generated by the perturber at the position $\bmath{R}_{\rm p}(t)$.
Like in  Paper~I, an artificial viscosity
has been introduced in the momentum equations. Further details about 
the numerical scheme adopted are given in the appendix.

We will also investigate the more general case in which
Eq.~(\ref{eq:fluid2}) is replaced by
\begin{equation}
\frac{\partial \bmath{v}}{\partial t}+(\bmath{v}\cdot
\bmath{\nabla})\bmath{v}=
-\frac{\bmath{\nabla}P}{\rho}-\bmath{\nabla}(\phi_{\rm G}+\phi_{\rm p})
+\mbox{viscous force},
\end{equation}
and the specific entropy, $s$, evolves according to
\begin{equation}
\frac{\partial s}{\partial t}+(\bmath{v}\cdot \bmath{\nabla}) s
=\mbox{viscous heating}.
\label{DsDt}
\end{equation}

The perturber, which is introduced instantaneously at $t=0$, moves in
the $(x,y)$-plane and evolves according to the equation
\begin{equation}
M_{\rm p}\frac{d^{2}\bmath{R}_{\rm p}}{dt^{2}}=-M_{\rm
p}\bmath{\nabla}\phi_{\rm G}+\bmath{F}_{\rm df},
\end{equation}
with the
dynamical friction force, $\bmath{F}_{\rm df}$, given by
\begin{equation}
\bmath{F}_{\rm df}=\int \delta\rho\,\bmath{\nabla}\phi_{\rm p}\,
d^{3}\!\bmath{r},
\label{dynfric}
\end{equation}
where $\delta\rho=\rho(\bmath{r},t)-\rho(\bmath{r},0)$. In all
the simulations presented here
$\bmath{V}\cdot\bmath{F}_{\rm df} \leq 0$ and $\bmath{F}_{\rm df}\cdot
\hat{\bmath{e}}_{\bmath r}\leq 0$
at any time of the run, where $\hat{\bmath{e}}_{\bmath{r}}$ is the unit
vector in
the radial direction.

A symmetry condition is applied at $z=0$. Apart from that
the domain is a rectangular box with open boundary conditions.
The size of the box must be taken large enough to ensure that the
density
perturbations that have propagated outside the domain
do not contribute significantly to the friction integral
(\ref{dynfric}).
To improve matters, additional spatial extend has been gained by
implementing a non-uniform mesh (see appendix for details).
Most of the simulations
were carried out with the grid represented in
Fig.~\ref{Nonuni}.

\begin{figure}
\epsfxsize=8.0cm\epsfbox{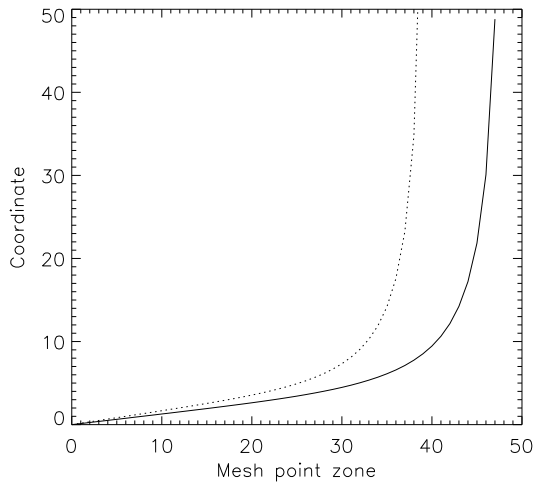}
  \caption{Coordinates for meshpoints in the non-uniform grid with
a transformation as given by Eq.~(\ref{eq:transf}). The dotted line
corresponds to the $z$-coordinate and the solid line for the $x$ and
$y$-coordinates.}
  \label{Nonuni}
\end{figure}

We choose units such that $G=R_{\rm G}=\rho(\bmath{0},0)=1$, where
$\bmath{r}=\bmath{0}$ corresponds to the centre of the primary and
$t=0$ is the beginning of the simulation.  The parameters to be
specified in the polytropic case are $\Gamma$, the isothermal sound
speed $c_{{\rm s}0}$, the mass of the central galaxy $M_{\rm G}$, the
mass $M_{\rm p}$ and softening radius of the satellite, and its initial
distance to the centre, and velocity. We do not try to simulate
the complex case in which the accretion radius is larger than
the softening radius. That would require a detailed stability analysis
of the flow. Objects with softening
radius larger than their accretion radius, like galaxies in clusters,
might only suffer from flip-flop instabilities if a large
amount of gas is replenished \cite{Bal94}. It is still unclear
whether this instability occurs in 3D or not.

Two requirements are imposed in
the initial distance of the satellite.  Firstly, the initial distance
is required to be a few galactic core radii ($R_{\rm G}$) and,
secondly, we will consider orbits such that $v_{\rm c}(R_{\rm
apo})/c_{\rm s}(R_{\rm apo})>1$, where $R_{\rm apo}$ is the apocentric
distance of the perturber's orbit in the absence of drag, $v_{\rm
c}(r)=\sqrt{rd\phi_{G}/dr}$ the circular velocity and $c_{\rm
s}^{2}\equiv \Gamma P/\rho$ for a polytrope.  The first condition is
necessary in order to have the complete history of the evolution of the
merging of an `external' satellite. The second condition comes from the
virial theorem as follows. Applying the virial theorem for a spherical
system with no fluid motions we get
\begin{equation}
{\mathcal{V}}+3(\Gamma-1){\mathcal{U}}-4\pi r_{S}^{3}
P\bigg|_{r_{S}}=0,
\end{equation}
where ${\mathcal{V}}=-\int_{0}^{r_{S}}\rho v_{\rm
c}^{2}\,d^{3}\bmath{r}$ and
$\mathcal{U}$ is the total internal energy within the sphere of radius
$r_{S}$. We
immediately obtain that
\begin{equation}
\frac{v_{\rm c}^{2}}{c_{\rm s}^{2}}=-\frac{r}{\rho}\frac{\partial
\rho}{\partial r}
\end{equation}
for a polytrope, and
\begin{equation}
\frac{v_{\rm c}^{2}}{c_{\rm s}^{2}}=-\frac{r}{\gamma\rho}\frac{\partial
\rho}{\partial r}
\end{equation}
for an isothermal background, assuming that perturbations are
adiabatic. Since the density should decay like $1/r^{2}$ or faster
($r^{-n}$ with $n\geq 2$) in the outer radii, the Mach number satisfies
${\mathcal{M}}=v_{\rm c}/c_{\rm s}\geq\sqrt{2/\gamma}$ in the outer
parts.  Therefore, for a monatomic gas, circular orbits in the outer
parts are always supersonic.

\section{Results}
\subsection{The reference model}

In this subsection the evolution of the perturber is discussed in a
reference model.  Variations from this model are considered in
subsequent subsections. Our reference model uses a resolution of
$95\times 95\times 40$ meshpoints, and a non-uniform grid as shown in
Fig.~\ref{Nonuni}. The gas is assumed to be a polytrope with
$\Gamma=1$.  The satellite has a mass $M_{\rm p}=0.3$, softening radius
$R_{\rm soft}=0.5$, and an initially circular orbit of radius $R_{\rm
p}(0)=4$.  We have chosen $M_{\rm G}=10$ and $c_{\rm s}=1.3$, so the
mass of gas within $R_{\rm p}(0)$ is approximately $M_{\rm G}$
\footnote{Typical parameters for galaxies are $R_{\rm G}=10$ kpc,
$\rho_{0}=10^{-24}$ g/cm$^{3}$, so the mass unit is $10^{10}$ M$_{\odot}$.}.  
Thus, the satellite mass is $\sim 1/30$ of the galaxy mass, and the
softening radius is three times the Bondi radius. Notice that
the perturber will move
supersonically until a radius of approximately $0.48$.

\begin{figure*}
\epsfxsize=18.0cm\epsfbox{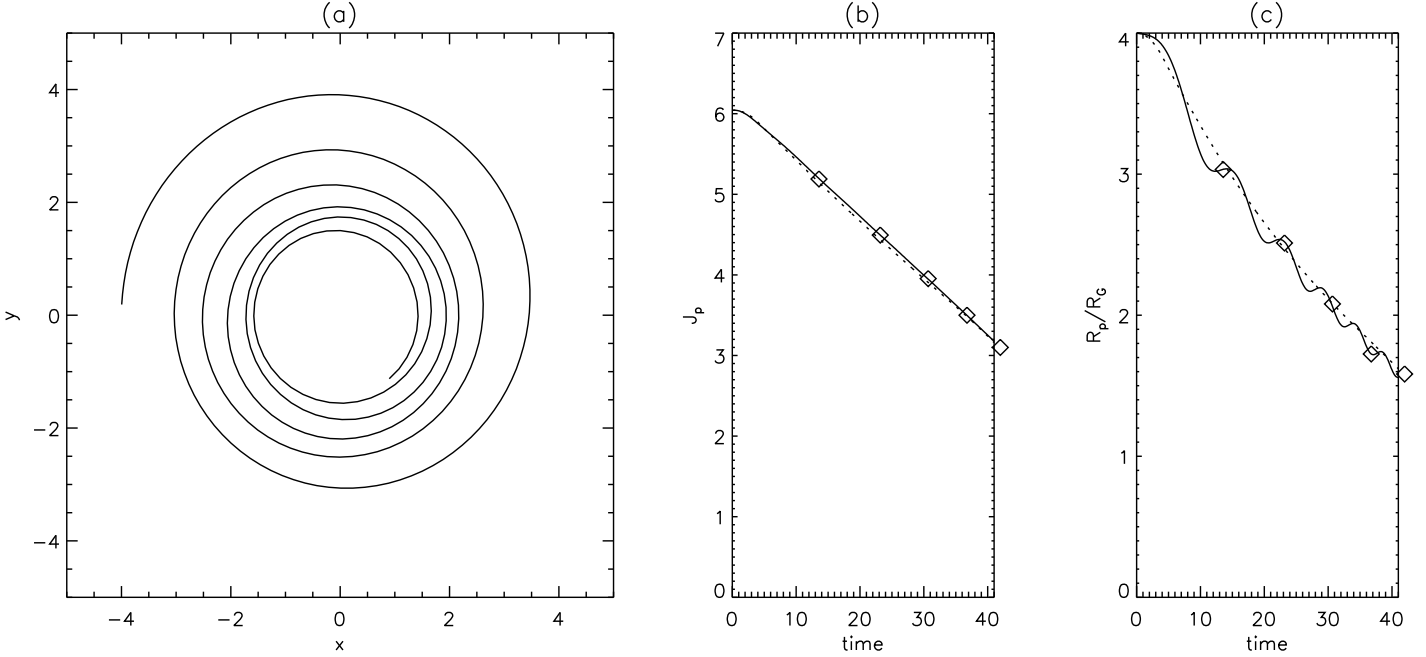}
  \caption{Evolution of the satellite in the reference model. The time
unit is
$\left(G\rho(\bmath{0},0)\right)^{-1/2}$. In panel (a) is plotted
the orbit in the $(x,y)$-plane. The angular momentum (per unit mass) and
orbital radius
of the perturber are shown as solid lines in panels (b) and (c),
respectively,
together with the predictions of the Local Approximation Prescription
(LAP; dotted lines). Symbols mark the time
when the orbital phase is zero, so the time between two consecutive
symbols is one
orbital period.}
  \label{standard}
\end{figure*}

\begin{figure*}
\epsfxsize=17cm\epsfbox{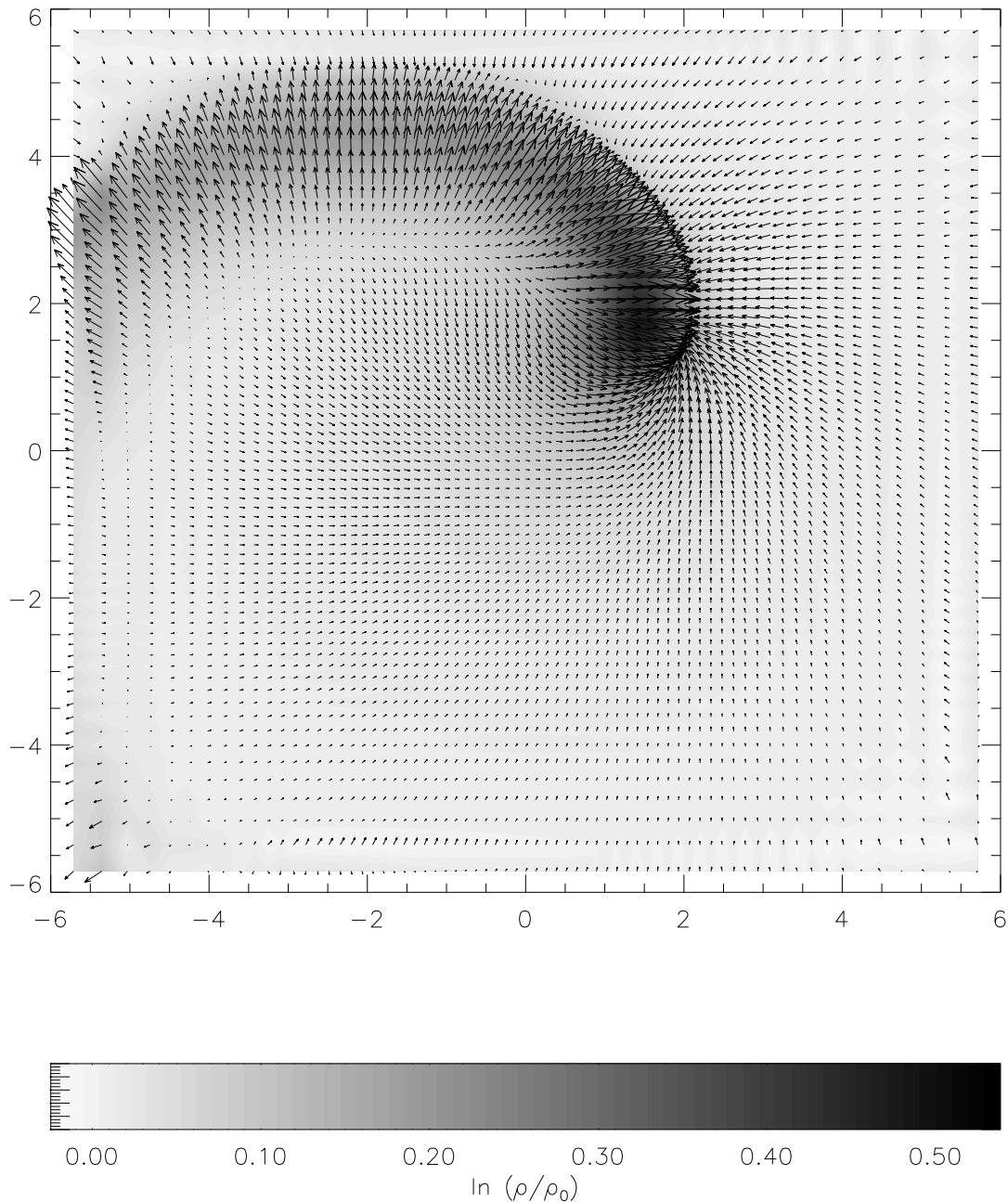}
  \caption{Grey-scale plot of the density enhancement
$\ln\left(\rho(\bmath{r},t)/
\rho(\bmath{r},0)\right)$, together with velocity vectors, at $z=0$ and
$t=26$, for the reference model with $M_{\rm p}=0.3$ and
$R_{\rm soft}=0.5$.}
  \label{Grey}
\end{figure*}

\begin{figure}
\epsfxsize=8.0cm\epsfbox{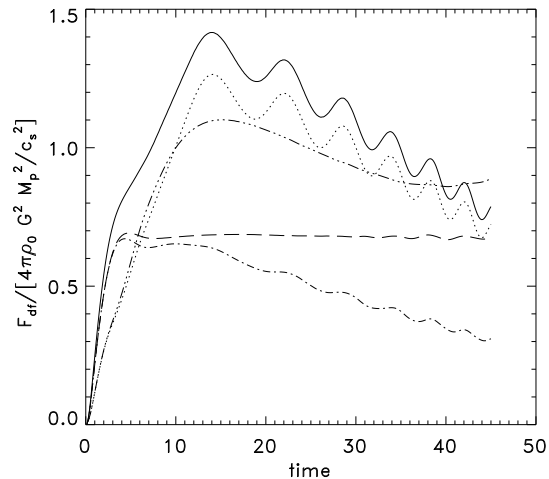}
  \caption{Different components and strength of $\bmath{F}_{\rm df}$
versus time
for the reference model. $F_{\rm df}$ (solid line), azimuthal component
of $\bmath{F}_{\rm df}$
(dotted line) and radial component of $\bmath{F}_{\rm df}$ (dot-dashed
line)
for a freely decaying perturber. We have taken the absolute values but
it is worthwhile
to recall that both $\bmath{F}_{\rm df}\cdot\bmath{V}$ and
$\bmath{F}_{\rm df}\cdot \hat{\bmath{e}}_{\bmath{r}}$
have negative values. The temporal evolution of $F_{{\rm df},r}$
(triple dot-dashed line) and $F_{{\rm df},\psi}$ (dashed line) for
a perturber forced to rotate in a circular orbit with $R_{\rm p}(0)=4$
are also plotted.}
  \label{forces}
\end{figure}

The satellite's angular momentum per unit mass, $J_{\rm p}$, and
orbital radius are shown in Fig.~\ref{standard} as a function of time.
We see that the angular momentum decreases almost linearly with time
(solid line).  A snapshot of the density enhancement and the velocity
field at the plane $z=0$ and $t=26$ is shown in Fig.~\ref{Grey}. The
strength of the radial and azimuthal components of $\bmath{F}_{\rm
df}$, in units of $4\pi G^{2}M^{2}_{\rm p}\rho_{0}/c_{\rm s}^{2}$ 
($\rho_{0}\equiv \rho(\bmath{r},0)$ hereafter) evaluated at
the instantaneous position of the satellite (the `dimensionless force'
hereafter), are presented in Fig.~\ref{forces} for the standard model.
As expected, the dimensionless force saturates since the direction of
$\bmath{V}$ changes and, consequently, the wake behind the body
effectively `restarts'. The force shows some oscillations caused by the
combination of the interaction of the satellite with its own wake,
together with small epicyclic motions of the satellite. For a perturber
forced to rotate in its originally circular orbit at constant velocity,
the azimuthal force is constant with time (dashed line in
Fig.~\ref{forces}), reaching a plateau after half an orbital period.
Almost no oscillations in the force are seen in this case.

Let us compare the decay rate with the prediction of formulas
(\ref{eq:Ost1}) and
(\ref{eq:Ost2}). Denoting the azimuthal component of the force by
$F_{{\rm df},\psi}$,
this expression yields for $R_{\rm p}> 0.48$ (i.e.\ in the supersonic
regime):
\begin{equation}
\frac{dJ_{\rm p}}{dt}=-\frac{F_{{\rm df},\psi} R_{\rm p}}{M_{\rm
p}}=-\frac{4\pi G^{2}M_{\rm p}\rho_{0}
R_{\rm p}}{v_{\rm c}^{2}}I_{\rm adi},
\label{eq:r13}
\end{equation}
where $v_{\rm c}$ is the circular velocity,
and $I_{\rm adi}\equiv \ln\Lambda+\frac{1}{2}\ln
\left(1-{\mathcal{M}}^{-2}\right)$
a dimensionless quantity. From $J_{\rm p}=v_{\rm c}R_{\rm p}$ we have
\begin{equation}
\frac{dR_{\rm p}}{dt}=v_{\rm c}^{-1}\left[2-\frac{3}{2}\frac{R_{\rm
p}^{2}}{R_{\rm p}^{2}+1}
\right]^{-1}\frac{dJ_{\rm p}}{dt}.
\label{eq:r14}
\end{equation}
It is noted that now $\rho_{0}$ and ${\mathcal{M}}$ depend on the position.
In order to integrate Eq.~(\ref{eq:r14}), the only
remaining unknown is $\ln\Lambda$. Since $\ln\Lambda$ is no longer a
linear function
of time, we assume that $I_{\rm adi}$ takes the following
values for ${\mathcal{M}}>1$
\begin{equation}
I_{\rm adi}=\left\{ \begin{array}{ll}
\ln \left(\frac{v_{\rm c}t}{r_{\rm min}}\right)+\frac{1}{2}\ln
{\mathcal{S}}_{\mathcal{M}}
& \mbox{${\rm if} \;\;\;\; \tau<t<t_{\rm c}$} \\
{t\over\tau}\left[\ln \left(\frac{{\mathcal{M}}}{
{\mathcal{M}}-1}\right)+\frac{1}{2}\ln
{\mathcal{S}}_{\mathcal{M}}\right] &
\mbox{${\rm if} \;\;\;\;
t<\tau<t_{\rm c}$} \\
\ln\left(\frac{v_{\rm c}t_{\rm c}}{r_{\rm min}}\right)
+\frac{1}{2}\ln {\mathcal{S}}_{\mathcal{M}} &
\mbox{${\rm if} \;\;\;\;
\tau<t_{\rm c}<t$}
\end{array}
\right.
\label{eq:LAP}
\end{equation}
where
\begin{equation}
\tau\equiv{r_{\rm min}\over v_{\rm c}-c_{\rm s}},\quad
{\mathcal{S}}_{\mathcal{M}}\equiv 1-{\mathcal{M}}^{-2},\quad
t_{\rm c}\equiv{\beta R_{\rm p}\over v_{\rm c}+c_{\rm s}},
\end{equation}
with $\beta$ being a free parameter which is chosen to match the values
of $J_{\rm p}(t)$ and $R_{\rm p}(t)$ obtained numerically. The
minimum
radius is taken to be $r_{\rm min}=2.25R_{\rm soft}$, as was
inferred for homogeneous media (Paper I) provided that
the softening radius is larger or equal to the accretion
radius. For
$t<\tau<t_{\rm c}$ a linear interpolation has been adopted.
For times such that $t_{\rm c}<\tau$,
it is difficult to suggest any prescription from the linear
theory. We therefore assume that the value remains approximately
unchanged compared
to the value it has got just before coming into that interval. However,
this assumption
is not completely satisfactory for bodies moving initially at Mach
numbers close to $1$.
Therefore we additionally assume that
\begin{equation}
I_{\rm adi}={t\over\tau}\left[\ln \left(\frac{{\mathcal{M}}}{
{\mathcal{M}}-1}\right)+\frac{1}{2}\ln
{\mathcal{S}}_{\mathcal{M}}\right]
\label{eq:LAP2}
\end{equation}
if $t<\tau$ and ${\mathcal{M}}>1$, regardless of the condition
$\tau< t_{\rm c}$.

For ${\mathcal{M}}<1$ and a homogeneous background the linear theory is
not able to capture the temporal evolution of the drag force but just
the asymptotic values; see Eq.~(\ref{eq:Ost2}). As a consequence, for a
body moving on a curved orbit, no reliable estimate of the drag force
can be proposed straightforward. In the following we shall refer to
Eq.~(\ref{eq:Ost2}) together with Eqs~(\ref{eq:LAP}),(\ref{eq:LAP2}) as
the Local Approximation Prescription (LAP).  It is `local' in the sense
that the force depends only on quantities at the instantaneous position
of the satellite even though, of course, the wake extends far away
beyond the satellite.  At this stage it becomes clear that the LAP may
fail at least for ${\mathcal{M}}$ close to or less than unity.

\begin{figure}
\epsfxsize=8.0cm\epsfbox{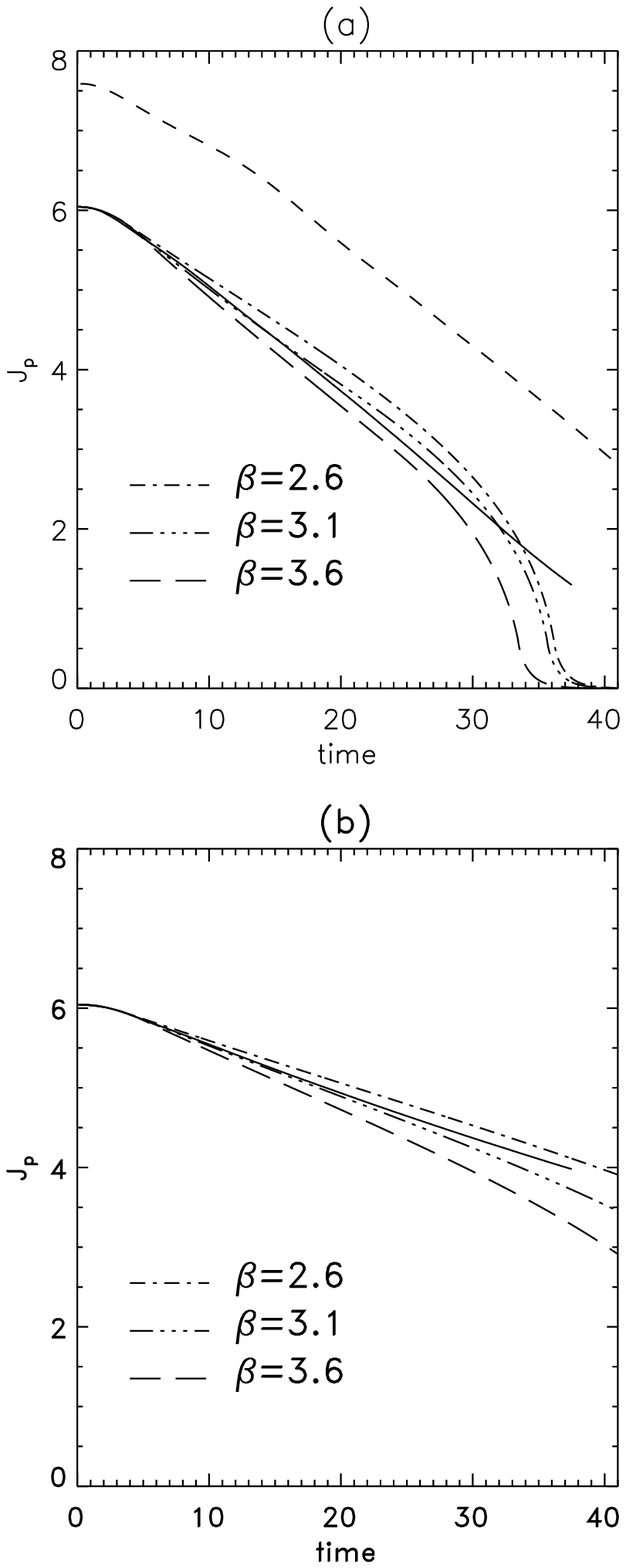}
  \caption{(a) Temporal evolution of the angular momentum (per unit mass)
for the isothermal model
with $M_{\rm p}=0.5$, $R_{\rm soft}=0.5$ and $R_{\rm p}(0)=4$ (solid
line) and for an
identical perturber but initially at $R_{\rm p}(0)=6$ (dashed line).
Panel (b) is like panel (a) but now $R_{\rm soft}=1$ and $R_{\rm
p}(0)=4$. The predictions by LAP with different $\beta$ are also
plotted for the perturbers initially at $R_{\rm p}(0)=4$. LAP overestimates
the drag when the distance of perturber to the centre becomes
comparable to $R_{\rm soft}$.}
  \label{ranging}
\end{figure}

The results of the integration for the reference model are plotted as
dotted lines in Fig.~\ref{standard} for $\beta=3.1$.  We see that it
agrees closely with the numerical values. Part of the success of Eqs
(\ref{eq:LAP})-(\ref{eq:LAP2}) resides in the fact that the condition
$\tau<t_{\rm c}$ is satisfied very soon, at $R_{\rm p}\simeq 3.5$, for
the present case.  However, it may be somewhat different for perturbers
moving with Mach numbers close to $1$ for a significant interval of
time. This case will be considered in Sect.~\ref{sec:poly}.

We ran a model which was identical to the reference model except that
the perturber mass was increased to $M_{\rm p}=0.5$. The results of the
simulation are plotted in Fig.~\ref{ranging}a, together
with LAP curves of different $\beta$. There is very good
agreement with the LAP for $\beta=3.1$. In general, it is found that a
change of a factor three in the mass of the perturber produces a slight
variation of about five per cent in the azimuthal component of the
dimensionless drag force.  Fig.~\ref{ranging}a also shows $J_{\rm p}(t)$ for
the same perturber but initially at orbital radius $R_{\rm p}=6.0$.
Once the angular momentum has decayed to $6.0$ the further evolution is
quite similar to the evolution of $J_{\rm p}$ starting at $6.0$.

A new run was done for $M_{\rm p}=0.5$ and $R_{\rm soft}=1$. The
evolution of $J_{\rm p}$ is plotted in Fig~\ref{ranging}b. Again the
LAP using $\beta=3.1$ reproduces reasonably well its evolution.

For simulations with resolutions 
$120\times 120\times 54$ the change in the drag force
is less than 0.5 per cent.

A measure of the robustness of the fitting formulae is given by the
dispersion
of $\beta$ for forced circular orbits of different radii, which will
be referred to as $\beta_{\rm cir}$. Using for $\beta_{\rm cir}$ the
expression
\begin{equation}
\beta_{\rm cir}=\frac{r_{\rm min}}{R_{\rm p}}
\frac{1+{\mathcal{M}}^{-1}}{(1-{\mathcal{M}}^{-2})^{1/2}}\exp\left(I_{\rm
adi}\right),
\label{eq:beta_def}
\end{equation}
with $I_{\rm adi}$ being the asymptotic value of the azimuthal
component of the force obtained numerically, we have $\beta_{\rm
cir}=2.6,2.9,3.2$ and $3.5$ for circular orbits of radii $R_{\rm
p}=4,3,2$ and $1.5$, respectively. The rotation velocities were taken
according to the rotation curve of the Plummer model. If the perturber
moves on a circular orbit of radius $R_{\rm p}=4$ but
${\mathcal{M}}=1.61$, $\beta_{\rm cir}$ is found to be $3.6$ (see Table
1). These results suggest that $\beta_{\rm cir}$ depends on
${\mathcal{M}}$, but could also contain information about the global
density profile.  In the inner regions $\beta_{\rm cir}$ should also
contain the effect of the excitation of gravity waves. Strong
resonances occur when the local Brunt-V\"{a}is\"{a}l\"{a} frequency
matches the orbital frequency of the perturber. The existence of
resonances depends strongly on the temperature profile in the inner
regions and require in general a large-scale entropy gradient in that
region (Balbus \& Soker 1990; Lufkin et al.\ 1993).  For satellite
galaxies the amplification of gravity waves, if any, may only take
place in the very inner part of the parent galaxy. However, other
effects, such as mass stripping and tidal deformation of the satellite
galaxy as well as friction with the luminous and collisionless parts of
the galaxy, may play an important role as far as dynamical evolution of
the satellite galaxy is concerned.

For fixed rotation curve, the parameter $\beta$ may have some
dependence on other variables of the system such as $\Gamma$. This
question will be considered in the next few sections.

\subsection{Testing the accuracy. Constant-velocity perturbers in a
homogeneous background}

A measure of the accuracy of our 3D simulations is given by comparing
the force of dynamical friction experienced by a perturber moving on a
straight-line orbit with constant velocity $\bmath{V}$ in a uniform and
isothermal medium, with the axisymmetric simulations described in Paper
I, which had higher resolution. In Fig.~\ref{2D3D} we plot the drag
force experienced by a body that moves in the $x$-direction using the
same resolution as that in our reference model, together with the
values obtained in high-resolution simulations.  The initial position
of the body is $(-4,-4,0)$ and it travels at constant velocity with
Mach number $0.75$ (bottom lines) and $1.12$ (upper lines).  In the
supersonic experiments the ratio between $R_{\rm soft}$ and the
accretion radius defined as $2GM_{\rm p}/V^{2}$, was $1.5$, with a
softening radius of $0.5$.  As long as the body is within the
region $|x|<6$ the accuracy of the force is better than five per cent.

\begin{figure}
\epsfxsize=8.0cm\epsfbox{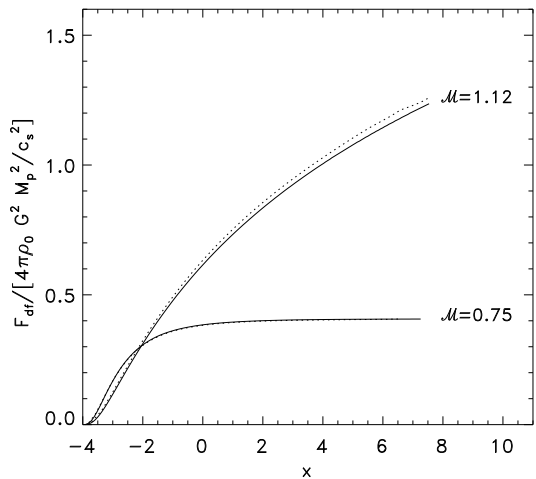}
  \caption{Dimensionless drag force experienced by a body moving
according to
$\bmath{R}_{\rm p}=(-4+Vt,-4,0)$ in a unperturbed background of
constant density. The solid line shows the numerical results
for high resolution axisymmetric simulations whereas the dotted line is
for
3D simulations with resolution as in the reference model.}
  \label{2D3D}
\end{figure}

\begin{figure}
\epsfxsize=8.0cm\epsfbox{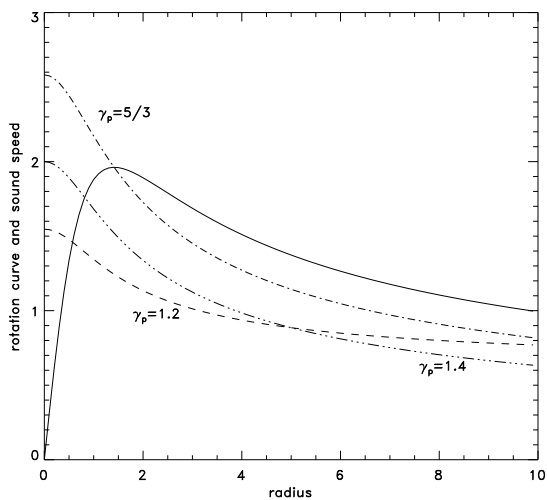}
  \caption{Rotation curve for the Plummer model (solid line), and the
sound
speed profiles for $\Gamma=5/3$ (dot-dashed line), $\Gamma=1.4$
(triple dot-dashed line), and $\Gamma=1.2$ (dashed line).}
  \label{Plummer}
\end{figure}

\subsection{Polytropic gas}
\label{sec:poly}
Let us now consider the same perturber with $R_{\rm soft}=0.5$ and mass
$M_{\rm p}$ that is forced to move on a circular orbit in a polytropic
gas and compute the value of $\beta_{\rm cir}$ that matches the
results. We do this for models which have different polytropic indices
$\Gamma$. The rotation curve is assumed to be the same as in the
reference model. Rotation curve and sound speed profiles for models
with different pairs $(\Gamma, c_{\rm s0})$ are drawn in
Fig.~\ref{Plummer}.  Since they have different polytropic indices we will
specify each of these models by just giving their polytropic index.  For
the model with polytropic index $5/3$, the Mach number of the perturber
at $R_{\rm p}=4$ is approximately the same as in the reference model.
The corresponding values of $\beta_{\rm cir}$ for each value of
$\Gamma$ are given in Table 1, except for the case $\Gamma=5/3$ at
$R_{\rm p}=2$ because for that orbit $t_{\rm c}\ll\tau$.  Table 1
suggests a remarkable dependence of $\beta_{\rm cir}$ on
${\mathcal{M}}$, as well as on $\Gamma$. This reflects the fact that
the asymptotic value of the azimuthal component of the dimensionless
force is not sensitive to the Mach number whenever
${\mathcal{M}}=1.1$--$1.6$. Given $\Gamma$ and $R_{\rm p}$, the
variation of the force in that interval of ${\mathcal{M}}$ is less than
five per cent.

\begin{table}
\caption{$\beta_{\rm cir}$ as defined in Eq.~(\ref{eq:beta_def}) for
models
which have different $\Gamma$. The body has a mass of $0.3$,
softening radius $0.5$, and is forced to move in a
circular orbit of radius $R_{\rm p}$ with a Mach number which may not
correspond to the rotation curve of the Plummer model. If $\beta_{\rm
cir}$
are displayed in the same row
for both $R_{\rm p}=4$ and $R_{\rm p}=2$, it means
that they correspond to the rotation curve given by the model.}

\begin{tabular} {c c c c c} \hline

{\bf $\Gamma$}
& {\bf ${\mathcal{M}}$ }
& {\bf $\beta_{\rm cir}$ } & {\bf ${\mathcal{M}}$ } & {\bf $\beta_{\rm
cir}$ }  \\

 &  $R_{\rm p}=4$ &  $R_{\rm p}=4$ &  $R_{\rm p}=2$ &  $R_{\rm p}=2$ \\
\hline \hline
1.0 & 1.16 & 2.6 & 1.45 & 3.2 \\
1.0 & 1.61 & 3.6 &      &     \\
1.2 & 1.61 & 2.8 & 1.67 & 3.1 \\
1.2 & 1.27 & 2.1  &     &  \\
1.2 &      &      &1.41 & 2.7 \\
1.4 & 1.27 & 2.2 & 1.41 & 2.7  \\
1.4 & 1.61 & 3.0 &      &     \\
1.4 &      &     & 1.67 & 3.0 \\
5/3 & 1.12 & 2.4 & 1.09 & ...  \\
\hline \hline

\end{tabular}
\end{table}

For sinking orbits the value of $\beta$ may differ somewhat from the
median of $\beta_{\rm cir}$ because the orbit of the body loses
correlation
with its wake faster than in a fixed circular orbit.
The case of $\Gamma=5/3$ is particularly interesting because then
$t_{\rm c}<\tau$ for a remarkably broad range of $R_{\rm p}$.
The $\beta$-parameter seems to be rather arbitrary in that range.
In Fig.~\ref{Gamma53} the evolution of $J_{\rm p}$ and $R_{\rm p}$ is
plotted for $M_{\rm p}=0.3$
(upper panel) and $M_{\rm p}=0.1$ (bottom panel). In both simulations
$\Gamma=5/3$.
The prediction of the LAP is plotted as dotted lines for $\beta=2.1$
in both cases.
In the first case the decay is
so fast that the LAP might no longer be a good approximation (upper
panel). However, even
for slower decays (the case of $M_{\rm p}=0.1$) the LAP
is not able to give reliable values of the force in the subsonic regime.
In fact, Eq.~(\ref{eq:Ost2}) overestimates significantly
the friction, mainly because of two reasons;
firstly, the wake must restart as a consequence of the curvature of the
orbit
and, secondly, the orbital radius and size of the object become
comparable.
\begin{figure}
\epsfxsize=8.0cm\epsfbox{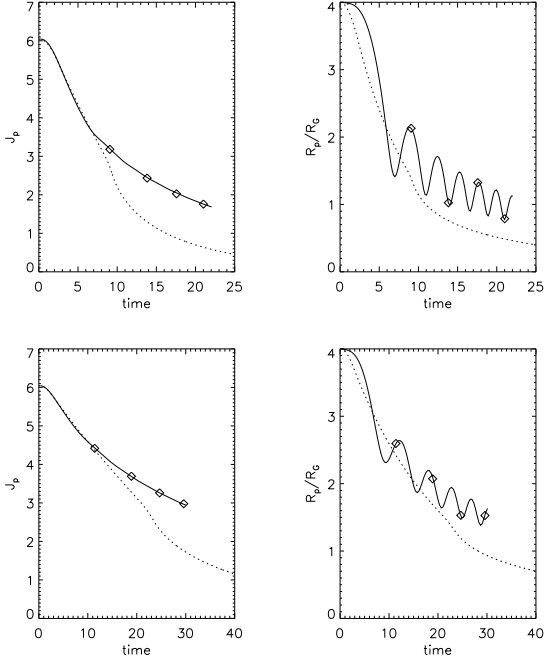}
  \caption{In the left panels it is shown the evolution of the
angular momentum (per unit mass)
for the model with $\Gamma=5/3$ and $M_{\rm p}=0.3$ (solid line of upper
panel)
and $M_{\rm p}=0.1$ (bottom panel). The corresponding evolution of the
radial distance
of the perturber is plotted as solid lines in the right panels. Dotted
lines show the predictions of the LAP using $\beta=2.1$. LAP fails
as soon as the motion becomes slightly supersonic.}
  \label{Gamma53}
\end{figure}
\begin{figure}
\epsfxsize=8.0cm\epsfbox{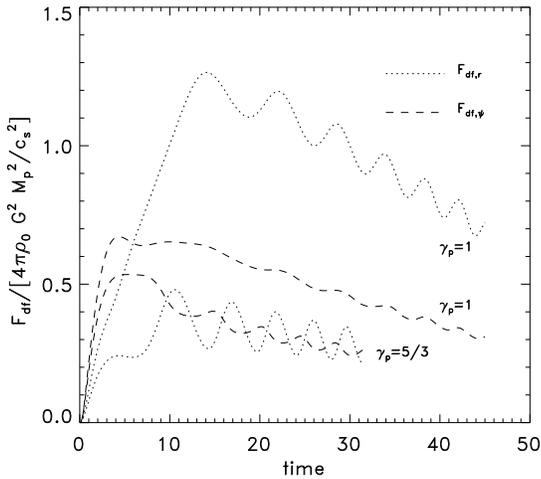}
  \caption{Solid lines show the value of the radial component of the
force $\bmath{F}_{\rm df}$
for the $\Gamma=1$ model with $M_{\rm p}=0.3$ (at the top) and for
$\Gamma=5/3$ and
$M_{\rm p}=0.1$ (at the bottom). The
azimuthal components are plotted with dashed-lines, whilst the radial
ones in dotted lines.}
  \label{force153}
\end{figure}

Since for the model with $\Gamma=5/3$ the typical
velocities $v_{\rm c}/c_{\rm s}$ lie in the range $1.0$--$1.2$ at the
radius interval $\in [2,4]$,
one would expect, according to Eq.~(\ref{eq:Ost1}),
a larger value of the dimensionless drag force in this case compared to
the isothermal gas model in which the body moves with higher values of
${\mathcal{M}}$.
A comparison between those forces for $\Gamma=5/3$ and $\Gamma=1$
is given in Fig.~\ref{force153}. Unexpectedly, the dimensionless drag
force is stronger
in the run corresponding to $\Gamma=1$.

Finally, the evolution of $J_{\rm p}$ of a perturber of mass $0.3$
is drawn in Fig.~\ref{Gamma1214} for $\Gamma=1.4$
as well as for $\Gamma=1.2$. The $\beta$-parameter was $2.2$ and $3.1$,
respectively.
Clearly, the LAP becomes inaccurate as soon as the body becomes
transonic.
\begin{figure}
\epsfxsize=8.0cm\epsfbox{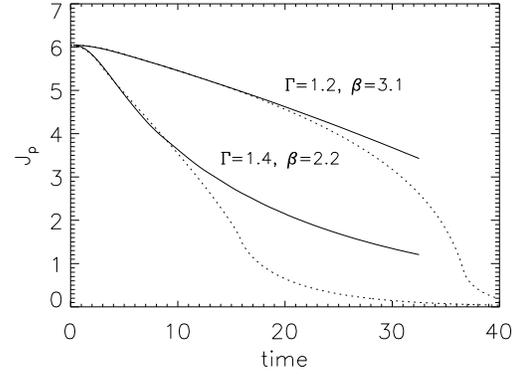}
  \caption{Decay of $J_{\rm p}$ of a perturber of mass
$M_{\rm p}=0.3$ initially in circular orbit, for different values of
$\Gamma$. The dotted lines indicate the prediction of the LAP.}
  \label{Gamma1214}
\end{figure}

\subsection{Non-circular orbits in the Plummer model}
The dependence of the orbital sinking times on the eccentricity of the
orbit is of interest for the statistical analysis of merging rates of
substructure in the cosmological scenario (e.g. Lacey \& Cole 1993;
Colpi et al.\ 1999), and for the theoretical eccentricity distributions
of globular clusters and galactic satellites \cite{Bos99}. On the one
hand, the hierarchical clustering model predicts that most of the
satellite's orbits have eccentricities between $0.6$ and $0.8$ (Ghigna
et al.\ 1998). On the other hand, the median value of the eccentricity
of an isotropic distribution is typically $0.6$ \cite{Bos99}. Here we
explore the dependence on eccentricity for dynamical friction in a
gaseous sphere in the Plummer potential. The non-singular isothermal
sphere is considered in the next subsection.

\begin{figure}
\epsfxsize=8.0cm\epsfbox{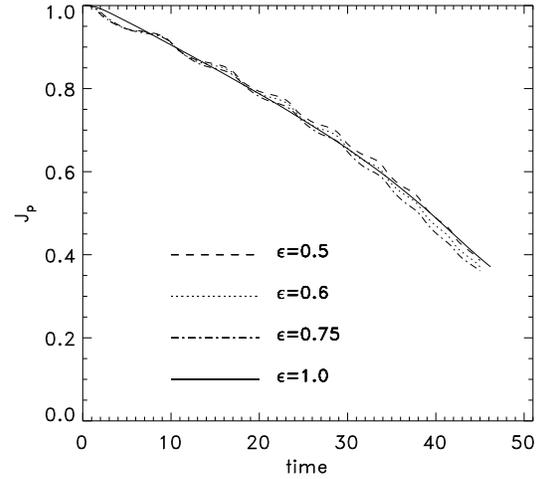}
  \caption{Panel (a) shows the orbital angular momentum
in units of the initial value for orbits having initially
the same energy but different eccentricities in the
Plummer model with $\Gamma=1$. $M_{\rm p}=0.4$ and $R_{\rm soft}=0.5$
were taken. In panels (b),
(c) and (d) the corresponding orbits in the plane $(x,y)$ are presented
for initial circularities $0.75$, $0.6$ and $0.5$,
respectively.}
  \label{eccPlummer}
\end{figure}

We shall discuss the sinking rate in terms of the angular momentum
half-time, $t_{1/2}$, the time after which the perturber has lost half
of its initial orbital angular momentum. Fig.~\ref{eccPlummer}  shows
the dynamical evolution of the satellite on bound orbits having
initially equal energies but different eccentricities ($J_{\rm p}$ is
in units of the initial value). Here we have used the polytropic model with
$\Gamma=1$.  A good fit to the numerical results is $t_{1/2}\propto
\epsilon^{-0.4}$, where $\epsilon$ is the initial circularity,
$\epsilon\equiv J_{\rm p}(E)/J_{\rm p; cir}(E)$, the ratio between the
orbital angular momentum and that of the circular orbit having the same
energy $E$. The dynamical friction time scale is found to increase with
increasing eccentricity. This finding makes a clear distinction between
dynamical friction in stellar and gaseous systems.  In fact, the
dynamical friction time scale has been obtained by direct N-body
simulations and in the linear response theory suggesting
$t_{1/2}\propto \epsilon^{0.4}$ for bodies embedded in a truncated
non-singular isothermal sphere of collisionless matter (van den Bosch
et al.\ 1999; Colpi et al.\ 1999).  A notorious difference between
stellar and gaseous backgrounds is that the gaseous force is strongly
suppressed when the perturber falls in the subsonic regime along its
non-circular orbit.  In order to discern how much the above result
depends on the model, we will consider in the next subsection the
angular momentum half-time for bodies orbiting in the non-singular
isothermal sphere.

\subsection{Sinking perturbers in the non-singular isothermal sphere}
\label{sec:sppis}
The flat behaviour of the rotation curves of spiral galaxies strongly
suggests the existence of isothermal dark halos around them. That is
the reason why most of the studies about dynamical friction assume an
isothermal sphere as the standard background. In the case of a
self-gravitating sphere of gas, the non-singular isothermal sphere is
also a natural choice. For instance, we may consider the dynamical
friction of condensed objects in early stages of a star cluster or a
galaxy, in which most of the mass of the system is gas. The gas
component is expected to follow the non-singular isothermal sphere out
to great radii.

We use the King approximation to the inner portions of an isothermal
sphere, with
softening radius $R_{\rm G}$ and central density $\rho_{\rm
G}(\bmath{0})$, i.e.
\begin{equation}
\phi_{\rm G}=-4\pi G \rho_{\rm G}(\bmath{0}) R_{\rm
G}^{2}\frac{1}{x}\ln\left(x+\sqrt{1+x^{2}}\right),
\end{equation}
with $x=r/R_{\rm G}$ \cite{Bin87}.
Our units are still $G=R_{\rm G}=\rho(\bmath{0},0)=1$. Even though the
rotation curve adopted is very similar to that of the Plummer model,
the detailed evolution, such as the dependence of the decay rate on
eccentricity or the level of circularization of the orbits, may be
somewhat different. For the sake of brevity, we only present the
results for the case where the Euler and continuity equations were
solved together with the entropy equation.

\begin{figure}
\epsfxsize=8.0cm\epsfbox{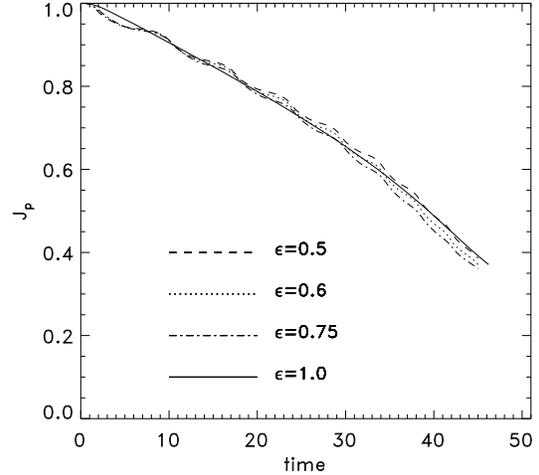}
  \caption{The orbital angular momentum
in units of the initial value for orbits having initially
the same energy but different circularities in the
King model.}
  \label{eccKing}
\end{figure}

Fig.~\ref{eccKing} shows the decay of a perturber with
$M_{\rm p}=0.3$ and $R_{\rm soft}=0.5$ in a King model with $\rho_{\rm
G}=1$, i.e.
pure gaseous sphere, for different circularities. There are
no appreciable variations of the decay time for different initial
circularities. In order to gain deeper insight
into the differences between the dynamical friction timescale
in an isothermal sphere of gas instead of collisionless matter,
we have computed the circularity at the times
when the orbital phase is zero (Fig.~\ref{circ}). The circularity
does not change at all for initially circular orbits. For eccentric
orbits, deviations in the circularity occur but there is no
net generation. The lack of dependence of the decay times on
circularity, together with the absence of any significant amount
of circularization, lead us to conclude that dynamical friction
in a gaseous isothermal sphere is not able to produce changes in the
distribution of orbital eccentricities. Mass stripping
which could lead to circularization has been ignored.

\begin{figure}
\epsfxsize=8.0cm\epsfbox{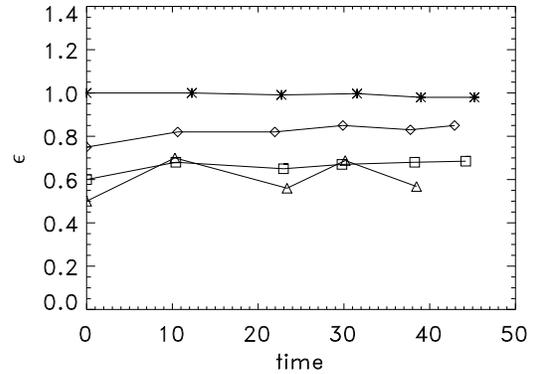}
  \caption{Circularity as a function of time for
the same situation as in Fig.~\ref{eccKing}.
The circularities are computed at the time when the
orbital phase is zero.}
  \label{circ}
\end{figure}

\section{Discussion and conclusions}

In this paper we have presented simulations of the decay of
a rigid perturber in a gaseous sphere. As a first
approximation, both the mass stripping and the barycentric
motion of the primary were neglected. In addition, the softening
radii of the perturbers were taken a few times larger than the
accretion radius. Simulations of the evolution of very compact
bodies ($R_{\rm soft}\ll R_{\rm ac}$), such as black holes
in a spherical background are numerically very expensive. 

The Linear Approximation Prescription is able to explain
successfully the evolution of the perturber provided the
(i) decay is not too fast, (ii) the motion is supersonic and (iii) the
body is not too close to the center to avoid that $R_{\rm p}$ becomes
comparable to $R_{\rm soft}$. In our simulations, the angular
momentum of the perturber decays almost linearly with time.
Generally speaking, the associated Coulomb logarithm
should depend on the distance of the perturber to the centre
and on its Mach number. However, the evolution
of a free perturber is relatively well described by just
a linear dependence of the maximum impact parameter of
the Coulomb logarithm on radius,
except very close to the centre where the time averaged
Mach number is less than unity.

In Chandrasekhar's formalism the Coulomb logarithm is often
approximated by the logarithm of the ratio of the
maximum to minimum impact parameters of the perturber.
In a singular isothermal sphere of collisionless matter, the time for
an approximately circular orbit to reach the center is
\begin{equation}
\tau_{\rm s}=1.2 \frac{r_{\rm cir}^{2} V_{\rm cir}}{G M_{\rm p}
\ln \frac{b_{\rm max}}{b_{\rm min}}},
\end{equation}
where $b_{\rm max}$ is roughly the virial radius of the
primary and $b_{\rm min}$ is of the order of the softening
radius of the perturber \cite{Bin87}. We have confirmed numerically
Ostriker's suggestion that the Coulomb logarithm should be replaced by
\begin{equation}
\ln \frac{b_{\rm max}}{b_{\rm min}}\rightarrow \frac{1}{0.428}
\ln \left(\eta\frac{R_{\rm p}(t)}{R_{\rm soft}}\right)
\end{equation}
for the gaseous case, with $\eta\simeq 0.35$, and independent of
the eccentricity. 
For very light perturbers
in near-circular orbits (objects with mass fractions smaller than one
per cent), the time scale for dynamical friction is a factor $2$
smaller in the case of a gaseous sphere than in corresponding stellar
systems (e.g.\ globular clusters in the Galactic halo). This factor is
somewhat smaller when $R_{\rm p}$ becomes comparable to a few $R_{\rm
soft}$. The weak dependence of the decay times on eccentricity suggests
that more eccentric orbits of merging satellites should not decay more
rapidly in the halo during the epoch in which matter is mainly gas.
Thus, they do not touch the disc at an earlier time than less eccentric
orbits.  Our simulations give support to the idea that dynamical
friction does not produce changes in the distribution of orbital
eccentricities.

The analysis must be modified if one wishes to account for the
existence of clouds in a non-uniform medium. So far, in a variety of
systems it is assumed that the perturber has condensed from
fragmentation of the gas; thus, the gaseous component must be thought
of as being distributed in clumps or clouds even more massive than the
`perturbers'.  In certain situations, compact perturbers may not be
dragged but heated (e.g.\ Gorti \& Bhall 1996). For this reason one may
be pessimistic towards a recent suggestion by Ostriker (1999) that,
because of the contribution of the gaseous dynamical friction force,
young stellar clusters should appear significantly more relaxed than
usually expected.

\section*{Acknowledgments}
We thank M.~Colpi and W.~Dobler for helpful discussions and
E.C.~Ostriker and M.~Ruffert for useful comments on the manuscript.
F.J.S.S. was supported by a TMR Marie Curie grant
from the European Commission. We acknowledge the hospitality of the
Institute for Theoretical Physics of the University of California,
Santa Barbara, where this work was finalized.

\appendix
\section{Description of the code}
We solve Eqs (\ref{eq:fluid1})--(\ref{DsDt}) on a nonuniform cartesian
mesh, $(x,y,z)$. This is accomplished using a coordinate transformation
to a uniform mesh, $(\xi,\eta,\zeta)$, with
\begin{equation}
x=x(\xi)={\xi\over1-(\xi/L_{\xi})^4},
\label{eq:transf}
\end{equation}
where $L_{\xi}$ is a length parameter,
and similarly for $y=y(\eta)$ and $z=z(\zeta)$. The transformed
equations are solved using sixth order centred finite differences,
\begin{equation}
\left({\partial^n f\over\partial\xi^n}\right)_i={1\over\Delta\xi^n}
\sum_{j=-3}^3 c_{j}^{(n)}f(x_{i+j})
\label{finitediff}
\end{equation}
for the $n$th $\xi$-derivative. The coefficients $c_{j}^{(n)}$ are given
in Table~\ref{Tcoef}. The expressions for the $\eta$ and $\zeta$
derivatives are analogous to Eq.~\ref{finitediff}.

\begin{table}\caption{Coefficients for the derivative formulae}
\begin{tabular}{ccccc}
\hline
$j$                    &    0   &  $\pm1$  & $\pm2$  & $\pm3$ \\ \hline
$c_{j}^{(1)}\times60$  &    0   & $\pm45$  & $\mp9$  & $\pm1$ \\
$c_{j}^{(2)}\times180$ & $-490$ & $+270$   &  $-27$  &  $+2$  \\ \hline
\end{tabular}
\label{Tcoef}
\end{table}

The corresponding $x$-derivative of a function $f(\xi(x))$ is obtained
using the chain rule, so
\begin{equation}
{\partial f\over\partial x}=
{\partial\xi\over\partial x}
{\partial f\over\partial\xi}={u'\over x'},
\end{equation}
where primes denote $\xi$-derivatives. For the
second derivative we have
\begin{equation}
{\partial^2 f\over\partial x^2}={u''x'-u'x''\over x'^3}.
\end{equation}
Again, the expressions for the $y$ and $z$ derivatives are analogous.
Since the scheme is accurate to $\Delta\xi^6$, $x$-derivatives are
accurate to $\Delta x^6(\xi)$, which varies of course across the mesh.

The third order Runge-Kutta scheme can be written in three steps
(Rogallo 1981):
\begin{equation}
\begin{array}{lll}
\mbox{1st step:}&
\quad f=f+\gamma_1\Delta t\dot{f},&
\quad g=f+\zeta_1\Delta t\dot{f},\\
\mbox{2nd step:}&
\quad f=g+\gamma_2\Delta t\dot{f},&
\quad g=f+\zeta_2\Delta t\dot{f},\\
\mbox{3rd step:}&
\quad f=g+\gamma_3\Delta t\dot{f},&
\end{array}
\end{equation}
where
\begin{equation}
\gamma_1={8\over15},\quad
\gamma_2={5\over12},\quad
\gamma_3={3\over4},\quad
\zeta_1=-{17\over60},\quad
\zeta_2=-{5\over12}.
\end{equation}
Here, $f$ and $g$ always refer to the current value (so the same space
in memory can be used), but $\dot{f}$ is evaluated only once at the
beginning of each of the three steps.
The length of the time step must always be a certain fraction of the
Courant-Friedrich-Levy condition, i.e.\ $\Delta t=k_{\rm CFL}\Delta
x/U_{\max}$, where $k_{\rm CFL}\leq1$ and $U_{\max}$ is the maximum
transport speed in the system.

The nonuniform mesh allows us to move the boundaries far away, so the
precise location of the boundaries should not matter. In all cases we
have used open boundary conditions by calculating derivatives on the
boundaries using a one-sided difference formula accurate to second
order. This condition proved very robust and satisfied our demands.

Equations (\ref{eq:fluid1})--(\ref{DsDt}) are solved in non-conservative
form. This is sufficiently accurate because of the use of high order
finite differences and because the solutions presented in this paper
are sufficiently well resolved.

\label{lastpage}
\end{document}